\documentclass[preprint2]{aastex}

\usepackage{amssymb}
\def\mnras{MNRAS}
\def\jcap{JCAP}

\def\apj{ApJ}

\def\apjs{ApJS}
\def\aj{AJ}

\def\prd{PRD}

\newcommand{\Mpc}{\mbox{Mpc}}

\newcommand{\SFH}{\mbox{SFH}}

\newcommand{\se}{\sigma}

\newcommand{\km}{\mbox{km}}
\newcommand{\s}{\mbox{s}}

\begin{document}

\title{Testing Homogeneity with Galaxy Star Formation Histories}
\author{Ben  Hoyle\altaffilmark{1},  Rita Tojeiro\altaffilmark{2}, Raul Jimenez\altaffilmark{3,1,4}, Alan Heavens\altaffilmark{5},  Chris Clarkson\altaffilmark{6}, Roy Maartens\altaffilmark{7,2}}
\altaffiltext{1}{Institut de Ciences del Cosmos (ICC), Universitat de Barcelona (IEEC-UB), Marti i Franques 1, Barcelona 08024, Spain.}
\altaffiltext{2}{Institute of Cosmology \& Gravitation, University of Portsmouth, Dennis Sciama Building, Portsmouth, PO1 3FX, UK.}
\altaffiltext{3}{ICREA, Institucio Catalana de Recerca i Estudis Avancats (www.icrea.es), Barcelona, Spain.}
\altaffiltext{4}{Theory Group, Physics Department, CERN, CH-1211, Geneva 23, Switzerland.}
\altaffiltext{5}{Imperial Centre for Inference and Cosmology, Astrophysics Group, Imperial College London, Blackett Laboratory, Prince Consort Road, London SW7 2AZ, U.K.}
\altaffiltext{6}{Astrophysics, Cosmology \& Gravity Centre, and, Dep. of Mathematics \& Applied Mathematics, University of Cape Town, South Africa. }
\altaffiltext{7}{Department of Physics, University of Western Cape, Cape Town 7535, South Africa.}

%
%

\begin{abstract}
Observationally confirming spatial homogeneity on sufficiently large cosmological scales is of importance to test one of the underpinning assumptions of cosmology, and is also imperative for correctly interpreting dark energy. A challenging aspect of this is that homogeneity must be probed inside our past lightcone, while observations take place on the lightcone. The star formation history (SFH) in the galaxy fossil record provides a novel way to do this.  We calculate the SFH of stacked Luminous Red Galaxy (LRG) spectra obtained from the Sloan Digital Sky Survey. We divide the LRG sample into 12 equal area contiguous sky patches and 10 redshift slices ($0.2<z<0.5$), which correspond to 120 blocks of volume $\sim$0.04\,Gpc$^3$. Using the SFH in a time period which samples the history of the Universe between look-back times 11.5 to 13.4 Gyrs as a proxy for homogeneity, we  calculate the posterior distribution for the excess large-scale variance due to inhomogeneity, and find that the most likely solution is no extra variance at all.  At 95\% credibility, there is no evidence of deviations larger than 5.8\%.\end{abstract}


\maketitle
\section{Introduction}
The $\Lambda$CDM concordance model is extremely successful, as it can fit most cosmological observations with just 6 free parameters  \citep{Komatsu:2010fb}. Testing the assumptions that go into this model is vital, but it is often neglected. In particular, the  model rests on the assumption of spatial homogeneity and isotropy on sufficiently large scales \citep[for a review see][]{2010CQGra..27l4008C,2011RSPTA.369.5115M,2012arXiv1204.5505C}. It is therefore appropriate and timely to devise observational tests that allow us to probe the homogeneity and isotropy assumptions. We know the isotropy assumption is well supported by detailed observations of the cosmic microwave background, which has shown that temperature variations are only one part in $10^5$ across the sky. However, homogeneity is much more difficult to probe.  Homogeneity is not established by observations of the CMB and the galaxy distribution: we cannot directly observe homogeneity, since we observe down our past light-cone, recording properties on 2-spheres of constant redshift and not on spatial surfaces that intersect that lightcone. What these observations can directly probe is isotropy. In order to link isotropy to homogeneity, we have to assume the Copernican Principle, i.e. that we are not at a special position in the Universe. The Copernican Principle is not observationally based; it is an expression of the intrinsic limitation of observing from one space-time location.

The importance of testing the homogeneity assumption has been highlighted by the development of inhomogeneous `void' models which can potentially explain apparent acceleration without any exotic physics. By changing the mean density and expansion rate radially away from us, observations such as SNIa can be accommodated without any dark energy \citep[see e.g.,][for reviews.]{2010JCAP...11..030B,2011CQGra..28p4004M,2012arXiv1204.5505C}. However, it is difficult to fit all observations~-- in particular the combination of $H_0$ and the CMB~-- without requiring significant inhomogeneity or other departures from the standard model at early times as well \citep{2011PhRvD..83f3506N,2012PhRvD..85b4002B,2011JCAP...02..013C,Roland2012}. 

This implies that tests for homogeneity must be made throughout the history of the universe. Consistency tests which could uncover deviations from homogeneity can be used to probe consistency of observables on our past lightcone \citep{2008PhRvL.101a1301C}.  Testing for the transition to homogeneity in the galaxy distribution on the lightcone, while assuming a Friedmann background, is another consistency test \citep[][]{2012MNRAS.425..116S}. Probing inside our past lightcone is harder, however, because we cannot observe it directly. One method is to use the Sunyaev-Zel'dovich effect to observe  CMB anisotropies from distant clusters \citep{1995PhRvD..52.1821G,2008PhRvL.100s1302C,2011RSPTA.369.5115M}. Another is to probe the thermal history in widely-separated  regions of the universe \citep{1986MNRAS.218..605B}, as it should of course be the same in the standard model. 

In this letter, we apply a new method of testing homogeneity in the interior of our past lightcone for the first time, by comparing the fossil record of galaxies at different redshifts at different times along their past world-lines, thus accessing different patches of the Universe at the same cosmic time.  A full proof of homogeneity would entail establishing homogeneity of the metric tensor. Here we apply a consistency test to check for violations of homogeneity, using the star formation rate as a probe, following the idea in \cite{2011JCAP...09..035H}. The fossil record, or the star formation history (hereafter $\SFH$), can be obtained by analysing  the shape of the galaxy spectrum, which encodes information about the histories of the component stellar populations, dust, and star formation. Various tools have been developed to extract this information \citep[e.g.,][]{Heavens:1999am,2005MNRAS.358..363C,2006MNRAS.365...46O,2012MNRAS.421.3116V}, of which we use the VErsatile SPectral Analysis\footnote{http://www-wfau.roe.ac.uk/vespa/} \citep[hereafter
VESPA, see][for more details]{Tojeiro:2007wt,Tojeiro:2009kk}. These approaches rely on the assumption that the evolution of the stellar populations is well understood and that the current modelling of stellar population is accurate. 
We use VESPA to obtain the $\SFH$ within the time bin 11.5 to 13.4 Gyrs of stacked Luminous Red Galaxy (LRG)  spectra located at different positions on the sky and at different redshifts. We compare the histories of different patches of the Universe, using the local star formation rate as a proxy for homogeneity.

This letter is organised as follows; in \S\ref{vespa2} we briefly describe the applicability of the VESPA routine as a test of homogeneity, and then describe the data, sstar formationimulated data and our method in \S\ref{data}. We present the results in \S\ref{results} and conclude in \S\ref{conclus}. 

To calculate distances, and to map from redshift to time, we assume a fiducial flat $\Lambda$CDM with best fit WMAP7 \citep{Komatsu:2010fb} cosmological parameter values ($\Omega_{\Lambda},\Omega_m,\Omega_b,\sigma_8,n_s,H_0$ = 0.729, 0.271, 0.045, 0.809, 0.966, 70.3 $\km\s^{-1}\Mpc^{-1}$). Since we are looking for deviations from homogeneity, it is conservative to assume this relation, which may be different in inhomogeneous universes \citep{2011JCAP...09..035H}. Any viable dark energy or modified gravity model will have
a background redshift-time relation that is close to the concordance
model's.

\section{VESPA and homogeneity}
\label{theory}
\label{vespa2}
\begin{figure}
   \centering
  \includegraphics[scale=0.275, clip=true, trim=9 4 4 4]{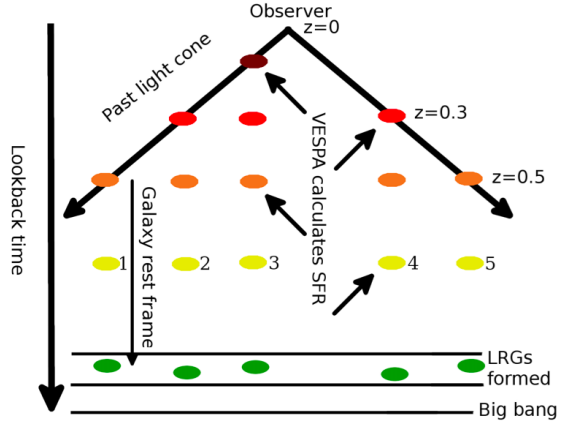}
   \caption{ \label{f0}  An illustration of the test of  homogeneity. We assume that LRGs form at a similar cosmic time, and have similar stellar
   formation rate histories (SFH), which we illustrate by the
   galaxies  changing colour. VESPA recovers the SFH for each galaxy
   in the galaxies rest-frame. We re-bin the SFH to the common lookback time, and compare the local star formation rate, for example, locations 1,2,..,5 to probe homogeneity.  Galaxy world lines are shown in comoving coordinates.}
\end{figure}

An illustrative diagram of our method is shown in Fig.\ref{f0}. Here, for illustration only, we assume that LRGs form at a similar cosmic time, and have similar $\SFH$s, which we illustrate by the galaxies changing colour. VESPA recovers the $\SFH$ for each galaxy
   along its own world line (vertical lines in  Fig.\ref{f0}), allowing us to compare the SFH at different distances but at the same cosmic time, e.g. at positions 1,2,..,5. 
In practice there is scatter in the SFH, due to sample variance on small scales and measurement error.  We will consider these later, and seek additional variance from large-scale inhomogeneity. 
\label{vespa}

\section{Data and Method}
\label{data}
\label{data}
\textbf{Data:} All of the galaxies used in this study were drawn from Sloan Digital
Sky Survey \citep[see][and references
therein]{York:2000gk,Gunn:2006tw,Smith:2002pca} Data Release 7
\citep[][hereafter SDSS DR7]{SDSSDR7}. We use $8.5\times10^4$
galaxies between the redshift range of $0.25<z<0.55$, selected to be Luminous
Red Galaxies \citep[][hereafter LRG]{Eisenstein:2001cq} drawn from the VESPA database.

We divide the SDSS survey footprint into $12$ equal area sky patches using
HealPix\footnote{http://healpix.jpl.nasa.gov} \citep[]{Gorski:2004by},
and $N_z=10$ redshift slices, whose widths are shown in Table \ref{t1}, together with  the  total number of galaxies, and the approximate volume of the SDSS survey in each redshift slice. We hereafter refer to the galaxies in each sky patch at each
redshift slice as a `block' ($B$) of galaxies. We randomly select galaxies in
each block into sub-samples of approximately 200 galaxies and 
stack the SDSS galaxy spectra for all galaxies in each sub-sample
following the method presented in \cite{Tojeiro:2010up}. 
Stacking the LRG spectra (as opposed to averaging the SFHs of individual galaxies in a block) allows us to recover the average star formation history of a block with higher resolution in lookback time \citep[see discussion in Sections 3.2 and 3.3 of][]{Tojeiro:2010up}. We use VESPA
to interpret the stacked spectrum in terms of a star formation and enforce VESPA
to recover measurements in 16 time bins, $\tau'$. The time bins are in
the rest-frame of the stacked spectra, or alternatively the rest-frame of
the galaxy block $T_B$, and we refer to these quantities as the Star
Formation Histories $\SFH(T_B,\tau')$. Additionally, we enforce VESPA to only allow star
formation in bins whose starting times are after the start of the
Universe, calculated assuming our fiducial cosmology.
\begin{table}
   \centering
  \begin{tabular}{| c | c | c | c |} 
  \hline
Redshift ID & Range & Ngals  & Total Vol Gpc$^3$\\ \hline
$1$ & $0.200<z<0.279$ & $7874$ & $0.90$\\
$2$ & $0.280<z<0.308$ & $9352$ &$0.46$\\
$3$ & $0.309<z<0.327$ & $8532$ &$0.34$\\
$4$ & $0.328<z<0.342$ & $8594$ &$0.29$\\
$5$ & $0.343<z<0.359$ & $9181$ &$0.36$\\
$6$ & $0.360<z<0.376$ & $8202$ &$0.39$\\
$7$ & $0.377<z<0.398$ & $8754$ &$0.55$\\
$8$ & $0.399<z<0.424$ & $8277$ &$0.71$\\
$9$ & $0.425<z<0.457$ & $8272$ &$1.00$\\
$10$ & $0.458<z<0.537$ & $8065$ &$2.91$\\
 \hline 
 \end{tabular}
    \caption{  \label{t1} \scriptsize{The redshift identifier and range of the
    redshift slices, the number of SDSS LRGs within each
    slice, and the approximate total volume in Gpc$^3$ contained by the redshift slice.}}
\end{table}

\begin{figure}
   \centering
  \includegraphics[scale=0.35, clip=true, trim=35 20 25 25]{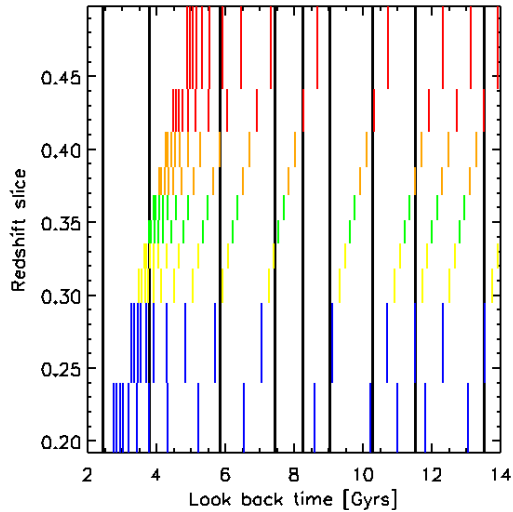}
   \caption{ \label{VESPAcommonFrame}  The start and end
positions of the rest-frame ($\tau'$ in the text) VESPA bins in lookback time, for each redshift slice. The continuous solid lines black lines show the
locations of the common frame ($\tau$) bins. We concentrate our test
of homogeneity within the final bin between 11.5 and 13.4 Gyrs.}
\end{figure}

\textbf{Methodology:}
\label{tdist_desc}
We next add the age of the Universe, calculated using our fiducial
cosmology, at the average redshift of the galaxy block $T_B$, to the ages of the
recovered VESPA bins for the stacked spectra. We map the values of the $\SFH(T_B,\tau')$ to a
common frame $\SFH(0,\tau)$ with bins denoted by $\tau$, the lookback time with respect to the current epoch,
and have chosen the lowest bin to
be at $t=0$. When we map the VESPA time bins $\tau'$ to the common
time bin $\tau$ we choose to maintain the bin widths, to avoid over-binning the
data.  Fig. \ref{VESPAcommonFrame} shows the start and end
times of the rest-frame $\tau'$ VESPA bins in lookback time (horizontal axis), for each redshift slice
(vertical axis). The continuous solid black lines show the
locations of the common frame $\tau$ bins with start times greater than 2 Gyrs.
LRGs form most of their stars very early on, and so we concentrate our
test of homogeneity within the greatest time bin $\tau=15$, which
corresponds to a look-back time between 11.5 and 13.4 Gyrs.  The reason for this is that all stacks have considerably (almost two orders of magnitude) more star formation in this time bin than others, and have smaller fractional errors (see below for the error assignment). The distribution of a large number of random variables (recall each stack has $\sim200$ LRGs spectra) can be modelled as a Gaussian, following the central limit theorem.  For lower $\tau<15$ bins some stacks have zero estimated star formation and larger fractional errors, skewing the distribution, which can no longer be modelled by a Gaussian, so we are unable to define a robust likelihood function.

The redshift slices of the blocks (recall that a block is a redshift slice/sky patch)
are chosen to contain $N_s\ge 3$ stacked spectra, which are constructed from the ($N_s$) galaxy sub-samples within the block. For each block $B$, we
calculate the average $A_B$, and estimate the standard deviation of the block SFH, $\se_B$,  from the sub-samples.
We determine the mean value $\mu$ of $A_B$, and further calculate the
average value of $A_B$ for all blocks at fixed redshift $z$, $A_z$  and the standard deviation of $A_z$
across the $N_z=10$ redshift slices, which we denote as $\sigma_z = \sigma \big( A_z\big).$

The dispersion $\sigma_z$ is scatter arising from the re-binning of solutions
$\SFH_{B,i}(T_B,\tau')$ of blocks at different redshifts
to the common frame $\SFH_{B,i}(0,\tau)$.  Note that block-to-block inhomogeneity would contribute to this, but only at the level of 1/12 of the variance, so it will affect our conclusions on the rms inhomogeneity by only 4\%.  We will, however, make this correction.

We compute the Student $t$-distribution $t_S$, for all blocks:
\begin{eqnarray}
\label{tdist}t_s &=& \frac{A_B - \mu}{\sqrt{\se_B^2 + \sigma_z^2} }\,,
\end{eqnarray}
which determines the number of `combined error' $\se \equiv \sqrt{\se_B^2 +
  \sigma_z^2}$, or the departure the measurement $A_B$, is from the
mean or notional value for the entire sample at each time $\tau$. 

For illustration we compare the probability density function obtained by the above
analysis, with the theoretical probability density function $f(t)$ of the t-distribution with $\eta=10\times12$ degrees of freedom, which has the analytic form given by 
\begin{eqnarray}
f(t) &=& \frac{1}{\sqrt{\eta} B(1/2,\eta/2)}\left( 1 + \frac{t^2}{\eta}
\right)^{-(\eta+1)/2}\,, 
\end{eqnarray}
where $B(1/2,\eta/2)$ is the Beta function.  We see in Fig. 3 that the
distribution of $t_s$ follows the expected distribution reasonably
well.  The grey shaded area shows the 95\% range for $t_s$ statistics
from 4000 Gaussian Random Samples of the SFH and errors.  

We now more formally model the data as having a gaussian distribution
but with the possibility of an extra fractional variance $V$ arising
from inhomogeneity.  i.e. we assume homogeneity and check for
consistency using the likelihood of the data given by
\begin{eqnarray}
\label{PVgauss}P_B(V)  \hspace{-3mm} &=&  \hspace{-3mm} \frac{1}{\sqrt{2\pi}\sigma_V} \exp{\left[-\left( A_B - \mu\right)^2/2\sigma_V^2\right]},\\
\sigma_V^2 \hspace{-3mm} &=& \hspace{-3mm} \se_B^2 + \sigma_z^2 + V\mu^2\,. 
\end{eqnarray}
If we assume a uniform prior for $V$, then $P(V) = \Pi_B P_B(V)$ is
the posterior for $V$ given the entire block dataset.  As a check, we
show $P(V)$ in Fig. 4 for simulated datasets (sub-blocks) with
variance $N_s \se_B^2$, for different star formation histories: a
continuous SFH, a gaussian SFH with mean 10 and standard deviation
$\sqrt{2}$ Gyr, an exponential SFH with a scale length of 0.5 Gyr, and
a SFH equal to the mean of the data.  SFHs are re-binned to the common frame.  We see in all cases that the posterior is correctly maximised at zero, and an upper limit dependent on the SFH.

\section{Results}
\label{results}
\label{results}
\begin{figure}
   \centering
  \includegraphics[scale=0.3, clip=true, trim=25 15 25 25]{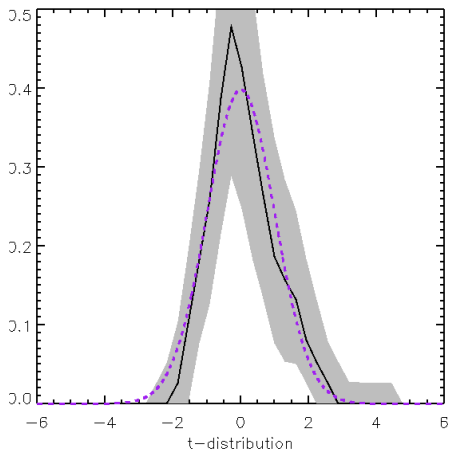}
   \caption{ \label{data2}  The Student-$t$ distribution for the SDSS SFH data, assuming that the variance is due to a combination of small-scale (sub-block) sample variance, measurement error and scatter due to rebinning from the galaxy rest-frame to the present epoch.  The vertical axis shows normalized frequency. The grey region give the 95\% spread of Gaussian Random Samples from the data and errors. We over-plot the theoretical $t_s$-distribution (dashed line).}
\end{figure}
 \begin{figure}[h!]
   \centering
   \includegraphics[scale=0.35, clip=true, trim=70 20 25 35]{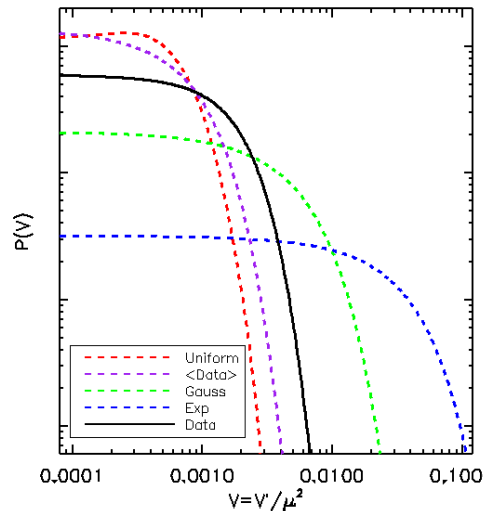}
   \caption{ \label{Pvdata}  The ensemble probability that the dispersion seen in the values of $A_B$ are drawn from a Gaussian distribution around $\mu$, as a function of an additional error component $V$, which is scaled by $\mu^2$. The colored dashed lines show the probability for each simulated SFH and the black solid line shows the data.}
\end{figure}

In Fig \ref{Pvdata} we show with the solid line the posterior distribution for the additional fractional variance $V$ in the SFH of the blocks.   The most probable variance due to inhomogeneity is zero, and 95\% of the posterior probability lies within $V<0.0032$.  Hence the 95\% credibility interval for the additional fractional inhomogeneity r.m.s., $\sqrt{V}$ (assuming a uniform prior on $V$), is 5.6\%, or 5.8\% if we include a correction for the rebinning. The colored lines show the different sets of simulated data. We see that only the Uniform SFH has a peak which is not at $V=0$, however it is consistent with 0 at the $<95\%$ confidence level. We note that by artificially reducing the additional variance on each block $\sigma_B$,  the peak moves closer to $V=0$.


\section{Conclusions}
\label{conclus}
\label{conclus}
Modern cosmology is built upon the assumption of homogeneity which is inferred through the observation of isotropy (e.g. by the Cosmic Microwave Background radiation) and the Copernican principle, stating that we do not occupy a preferred location.

Deviations from homogeneity, in particular, an inhomogeneous background, for example LTB models in which massive void exist, can potentially explain the dimming of distant supernovae without invoking dark energy. Testing homogeneity is therefore an active area of research, and many tests have been devised, e.g., kinematic SZ effect.

In this letter we have performed a new observational test of homogeneity \citep{2011JCAP...09..035H} by examining the estimated Star Formation Histories in old stars  from stacked spectra of SDSS Luminous Red Galaxies \citep[][]{Eisenstein:2001cq} using VESPA.  The data are blocks in 10 redshift intervals $0.025 < z < 0.55$, with 12 equal-area angular bins. 

We estimate the sample variance and measurement error arising from small-scale (sub-block) variations by computing the error on the mean of the sub-blocks.  Additionally, we include the scatter arising from re-binning to the present-day lookback time, and then perform a Bayesian analysis of any additional variance which may exist on large scales. Our test assumes homogeneity and checks for consistency and we find no evidence for extra variance, and a 95\% upper limit to the credibility interval of a fractional variation of 5.8\% in SFH between 11.5 and 13.4 Gyrs.  The typical block size is about 0.04 Gpc$^3$.  

The main uncertainty is in the stellar populations models employed by VESPA. However, this result can be easily extended and improved upon with future spectroscopic surveys e.g., BOSS \citep{2012arXiv1208.0022D}, and as our knowledge of Stellar Population Models increases.
Although this is not a complete test of homogeneity, which would require investigation of the metric tensor itself, this limit on homogeneity is the first to come from within the past light cone, rather than being restricted to our past light cone.  As such it is genuinely testing homogeneity rather than isotropy.

\section*{Acknowledgments} 
\label{ack}
BH  would like to thank Aday Robiana and Roland dePutter for useful discussions,  the University of Cape Town for hospitality, and acknowledges  grant number FP7-PEOPLE- 2007- 4-3-IRG n 20218.
CC and RM were supported by the South African NRF and by a UK Royal Society/ NRF exchange grant. RM was supported by the SA SKA Project and the UK STFC (grant ST/H002774/1).Funding
for the SDSS and SDSS-II has been provided by the Alfred
P. Sloan Foundation, the Participating Institutions, the
National Science Foundation, the U.S. Department of
Energy, the National Aeronautics and Space Administration,
the Japanese Monbukagakusho, the Max Planck
Society, and the Higher Education Funding Council for
England. The SDSS Web Site is http://www.sdss.org/.

\end{document}